\documentclass[twocolumn,aps,pra]{revtex4}
\usepackage{epsfig}
\usepackage[english]{babel}
\usepackage{latexsym}
\usepackage{graphics}
\usepackage{subfigure}
\usepackage{epsfig}
\usepackage{graphicx}
\usepackage{dcolumn}
\usepackage{amsmath}
\usepackage{hyperref}
\usepackage{amssymb}

\begin{document}

\title{Coulomb-induced ionization time lag after electrons tunnel out of a barrier}
\author{Y. J. Chen$^{1}$, X. J. Xie$^{1}$, C. Chen$^{1}$,  G. G. Xin$^{2,3}$, and J. Liu$^{4,5}$}

\date{\today}

\begin{abstract}

After electrons tunnel out of a laser-Coulomb-formed barrier, 
the movement of the tunneling electron can be  affected by the Coulomb tail.
 We show that this Coulomb effect induces a  large time difference (longer than a hundred attoseconds)
between the exiting time at which the electron exits the barrier
and the ionization time at which the electron is free. This large time difference
has important influences on strong-field processes such as above-threshold ionization and high-harmonic generation,
with remarkably changing time-frequency properties of electron trajectories.
Some semi-quantitative evaluations on these influences are addressed, which provide new insight into strong-field physics
and give important suggestions on attosecond  measurements.

\end{abstract}
\affiliation{1.College of Physics and Information Technology, Shaan'xi Normal University,Xi'an710119, China\\
2.School of Physics, Northwest University, Xi'an 710127, China\\
3.Shaanxi Key Laboratory for Theoretical Physics Frontiers, Northwest University, Xi'an710069, China\\
4.
Institute of Applied Physics and Computational Mathematics, Beijing 100088, China\\
5.CAPT, HEDPS, and IFSA Collaborative Innovation Center of MoE, Peking University, Beijing 100871, China
} 

\maketitle

\emph{Introduction}.---As  exposed to strong laser fields, the Coulomb potential of an atom is bent by the external field
with forming a barrier out of which the valence electron of the atom  can tunnel \cite{Keldysh,ADK}.
Thereafter,  the tunneling electron can be considered as a classical particle, which is moving in the external field
and can be driven back to and recollide with the parent ion.
These considerations are the cores of the classical two-step model (CM) \cite{Schafer1,Corkum},
which  is  successful in describing
many strong-field processes related to the tunneling electron,  such as above-threshold ionization (ATI) \cite{Kulander1,Kulander2,Lewenstein2},
high-harmonic generation (HHG) \cite{Soc,Schafer,Corkum1993,Lewenstein},
and non-sequential double ionization (NSDI) \cite{Hasbani2,Zeidler,Becker2}, etc..

Based on the CM, or  strong-field approximation (SFA) which can be considered as the quantum version of CM,
it is well established that there are long and short electron trajectories contributing to ATI \cite{Lewenstein2,Becker2002},
and similar situations also go for HHG \cite{Lewenstein}.
The long and short trajectories are well resolved in the temporal domain.
Specifically, for ATI, the long (short) electron trajectory has an ionization time located in the falling (rising)
part of the laser field in one laser cycle. In particular, the long trajectory is also associated with a rescattering process.
For HHG, both long and short electron trajectories are related to rescattering and have ionization times located in
the falling part of the laser field.
In comparison with the short one, the long trajectory has a earlier ionization time and a later return time.

Both the CM and the SFA which neglect the Coulomb effect are often used as benchmarks to deduce the electron attosecond dynamics \cite{Krausz,Vrakking}.
On the other hand, a great deal of studies have shown that the Coulomb potential has important influences on strong-field dynamics of
the tunneling electron \cite{Brabec,Milosevic2006}. For example, the Coulomb potential will remarkably affect
the ATI photoelectron energy spectrum \cite{Blaga}, momentum \cite{Eckle} and angular \cite{Goreslavski} distributions, etc..
Nevertheless, the influences of Coulomb potential on time-resolved strong-field electron dynamics have been unclear \cite{Shafir}.
Especially, quantitative even  semi-quantitative evaluations on these influences have been absent.
These evaluations are necessary for more precise attosecond measurements and controls of the electrons in atoms, molecules, and solids.
They are also important for establishing a full Coulomb-modified  physical picture of strong-field electron dynamics.

Here, we make efforts to clarify the Coulomb effect on
time-trajectory-resolved electron dynamics
for ATI and HHG. We begin our discussions with  ATI in an orthogonally polarized two-color (OTC) field, which allows one to
resolve the contributions of long and short electron trajectories directly from the photoelectron momentum distributions.
We show that the Coulomb potential induces a time difference of the tunneling electron
between the moments of exiting the barrier and  being free.
This time difference (longer than a hundred attoseconds)
is well mapped in the momentum distributions, resulting in a remarkable
increase of contributions of long trajectory (more than 20$\%$) to ATI. Then we extend our discussions to HHG.
We show that the exiting times of Coulomb-modulated HHG trajectories are earlier  (about 25 attoseconds) than expected,
leading to a marked increase of contributions of short trajectory
(about one order of magnitude) to HHG.

\emph{Numerical and analytical methods}.---
We assume that the fundamental field is along the x axis and  the additional second-harmonic field is along the y axis.
The Hamiltonian of  the model He atom studied here
has the form of H$(t)=H_0+\mathbf{r}\cdot \mathbf{E}(t)$ (in atomic units of $\hbar= e = m_{e} = 1$).
Here, the term $H_0=\mathbf{p}^2/2+V(\mathbf{r})$ is the field-free Hamiltonian, and $V(\mathbf{r})=-{Z e^{-\rho r^{2}}}/{\sqrt{r^{2}+\xi}}$ with
$r^{2}=x^{2}+y^{2}$ is the  Coulomb potential.
$\rho$ is the screening parameter with $\rho=0$ for the long-range potential and $\rho=0.5$ for the short-range one.
$\xi=0.5$ is the smoothing parameter, and $Z$ is the effective charge which is adjusted in such a manner that
the ionization potential of the model system reproduced here is $I_p=0.9$ a.u..
The term $\mathbf{E}(t)$ is the electric field of OTC which has the form of \cite{Kitzler,Zhang}
$\mathbf{E}(t)=\vec{\mathbf{e}}_{x}{E}_x(t)+\vec{\mathbf{e}}_{y}{E}_y(t)$
with ${E}_x(t)=f(t)E_{0}\sin{(\omega_{0}t)}$
and ${E}_y(t)=\mathcal{E} f(t)E_{0}\sin{(2\omega_{0}t+\phi)}$.
$\vec{\mathbf{e}}_{x}$  ($\vec{\mathbf{e}}_{y}$) is the unit vector along the  $x$ ($y$) axis.
$\phi$ is the relative phase between these two colors.
 ${E}_0$ is the maximal laser amplitude relating to the peak intensity $I$ of the fundamental field ${E}_x(t)$.
$\mathcal{E}$ is  the ratio of the maximal laser amplitude for the second-harmonic field $E_{y}(t)$ to  ${E}_0$.
$\omega_{0}$ is the laser frequency of $E_{x}(t)$ and $f(t)$ is the envelope function.
We use trapezoidally shaped laser pulses with a total duration of 20 optical cycles and linear ramps of three optical cycles.
The details for solving  TDSE of $i\dot{\Psi}(t)=$H$(t)\Psi(t)$ with spectral method \cite{Feit}
and obtaining the photoelectron momentum distribution can be found in \cite{Gao}.
Unless mentioned elsewhere, the laser parameters used are $I=5\times10^{14}$W/cm$^{2}$, $\omega_0=0.057$ a.u.,
$\phi=\pi/2$ and $\mathcal{E}=0.5$.

\begin{figure}[t]
\begin{center}
{\includegraphics[width=8.5cm,height=6.6cm]{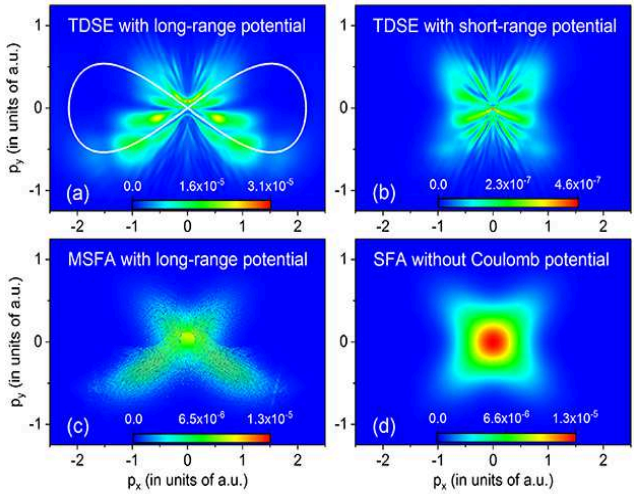}}
\caption{Photoelectron momentum distributions of He, obtained with TDSE simulations of
long-range (a)  and short-range (b) potentials, with analytical simulations of  MSFA that considers the effect of
a long-range potential (c) and  SFA where the Coulomb effect is omitted (d).
The prediction of  CM (the white line) is also plotted in (a).
}
\label{fig:graph1}
\end{center}
\end{figure}
\begin{figure}[t]
\begin{center}
{\includegraphics[width=8.5cm,height=6.6cm]{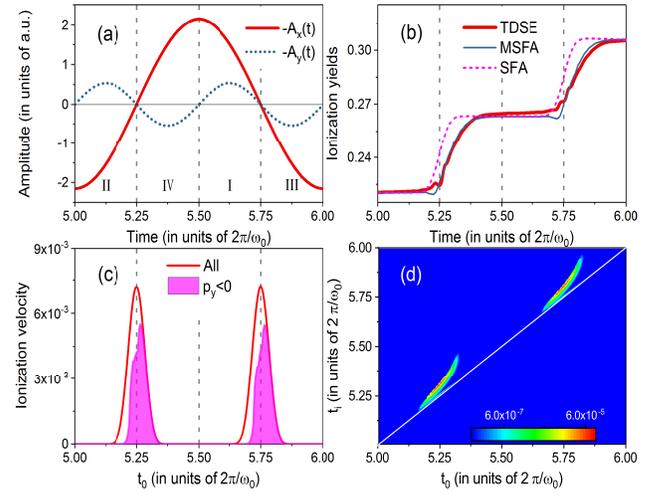}}
\caption{The prediction of  CM for the drift momenta of $p_x(t)=-A_x(t)$ and $p_y(t)=-A_y(t)$ (a),
the comparison of time-dependent ionization yields calculated with  TDSE,
 MSFA and  SFA (b), the ionization velocity predicted by  MSFA for all values of
$p_y$ and only for $p_y<0$ (c), the distribution as functions of the exiting time $t_0$ and the ionization time $t_{i}$ predicted by
MSFA (d), in one laser cycle  of $2\pi/\omega_0$.
For comparison,  MSFA and SFA curves in Fig. 2(b) are multiplied by a vertical scaling factor to match the TDSE one.
The log$_{10}$ scale is used in the distribution in (d).}
\label{fig:graph1}
\end{center}
\end{figure}

To analytically study the Coulomb effect on time-resolved dynamics of ATI, we first
calculate ATI electron trajectories, characterized by the complex ionization time $t_s=t_0+it_x$, the drift momentum $\mathbf{p}$
and the tunneling amplitude $F(\mathbf{p},t_s)\equiv F(\mathbf{p},t_0)\sim e^{b}$,
with SFA \cite{Lewenstein2,Becker2002}. Here, b is the imaginary
part of the quasiclassical action at relevant trajectories and
only minus values of b are considered.
Then as in \cite{Brabec,Goreslavski}, we solve the Newton equation
$\ddot{\mathbf{r}}(\mathbf{p},t)=-\mathbf{E}(t)+\nabla_\mathbf{r} V(\mathbf{r})$
for each SFA electron trajectory, with  initial conditions \cite{Beckeroribit} $\dot{\mathbf{r}}(\mathbf{p},t_0)=\mathbf{p}+\mathbf{A}(t_0)$
(the exiting momentum) and
$\mathbf{r}(\mathbf{p},t_0)=Re(\int^{t_0}_{t_s}[\mathbf{p}+\mathbf{A}(t')]dt')$ (the exiting position).
Here,  the real part $t_0$ of $t_s$  is considered as the exiting time,
and $\mathbf{A}(t)$ is the vector potential of $\mathbf{E}(t)$.
In our simulations, trajectories with $|\mathbf{r}(\mathbf{p},t)|\leq 4$ a.u. at $t>t_0$ are also absorbed.
The final Coulomb-modulated  drift momentum is obtained with $\mathbf{p}_f=\dot{\mathbf{r}}(\mathbf{p},t\rightarrow\infty)$,
which is relating to the amplitude $F(\mathbf{p},t_0)$.
Similarly, with finding the return time $t_r$ which satisfies the relation $\mathbf{r}(\mathbf{p},t_r)=0$ with $t_r>t_0$, we also
obtain the Coulomb-modulated HHG electron trajectories, characterized by the exiting time $t_0$, the return time $t_r$,
the return energy $E_p=[\dot{\mathbf{r}}(\mathbf{p},t_r)]^2/2$, and the amplitude $(1/\tau)^{1.5}F(\mathbf{p},t_0)$ with $\tau=t_r-t_0$.
This factor $(1/\tau)^{1.5}$ stands for quantum diffusion effects \cite{Lewenstein}.
Below, we will call the above Coulomb-modulated SFA  the MSFA.
Note, according to the SFA,
the exiting time $t_0$  agrees with the ionization time $t_i$
at which the value of $E_a(t_i)$ becomes larger than zero.
Here, the term $E_a(t)=[\dot{\mathbf{r}}(\mathbf{p},t)]^2/2+V(\mathbf{r})$ is the total instantaneous
energy of the tunneling electron.
However, as we will show in the following,  the MSFA which considers the Coulomb effect
predicts a time difference $t_{d}=t_i-t_0$ with $t_d>0$. This difference influences remarkably on dynamics of the laser-driven system.

\emph{Cases of ATI}.---
The calculated TDSE photoelectron momentum distributions for He with long-range and short-range potentials
are presented in  Figs. 1(a) and 1(b). The long-range TDSE results in Fig. 1(a) show a butterfly-like structure with
a remarkable up-down asymmetry with respective to the axis of $p_y=0$. The distribution in Fig. 1(a) has larger amplitudes for $p_y<0$.
By contrast, this up-down asymmetry remarkably decreases for the short-range results in Fig. 1(b).
Further simulations with SFA,  related to a delta potential, give an up-down
symmetric distribution, as shown in Fig. 1(d). Simulations with the MSFA  indeed reproduce
the remarkable up-down asymmetry, as shown in Fig. 1(c), implying that this asymmetry arises from the Coulomb effect.
The symmetry degree, which is defined as the ratio of the total amplitudes with $p_y>0$ to those of $p_y<0$,
is about 0.55  in Fig. 1(a) and 0.62  in Fig. 1(c).
The TDSE and MSFA results are near to each other. This degree is near the unity in Fig. 1(b) of short-range potentials.
For comparison, the prediction of the CM for $p_x$ versus $p_y$ is also plotted in Fig. 1(a),
which shows an up-down symmetric structure, similar to the SFA one.
We mention that we have  performed simulations  at various laser parameters.
On the whole, this TDSE symmetry degree is smaller for lower laser intensities or shorter laser wavelengthes,
in agreement with MSFA predictions.

A sketch of the CM prediction of drift momenta $p_x(t)=-A_x(t)$ and $p_y(t)=-A_y(t)$ for one laser cycle of the fundamental field
is presented in Fig. 2(a).
It clearly shows that with the steering of the second-harmonic field of OTC  \cite{Kitzler,Zhang}, the contributions of the long (short)
trajectory related to the fundamental field
are mapped in the third (III) and the fourth (IV) quadrants with $p_y<0$ (the first (I) and the second (II) quadrants with $p_y>0$),
providing a manner for resolving the contributions of long versus short electron trajectories directly from photoelectron momentum distributions.

To understand how the Coulomb effect influences the ATI, we further calculate the time-dependent ionization yields obtained by different methods.
The TDSE ones are obtained with evaluating $I(t)=1-\sum_{m}|\langle m|\Psi(t)\rangle|^2$.
Here, $|m\rangle$ is the bound eigenstate of H$_0$ obtained through imaginary-time propagation.
The model ones are obtained with calculating $I(t)=\sum_{\mathbf{p},t}|F(\mathbf{p},t_0)|^2$ at $E_a(t)>0$.
Relevant results are presented in Fig. 2(b). The SFA results in Fig. 2(b)
clearly show a remarkable increase just around the time of $t=5.25T$ or $t=5.75T$  with $T=2\pi/\omega_0$,
at which the fundamental field arrives at its peak.
The TDSE results, however, show a remarkable increase around a time later than $t=5.25T$ or $t=5.75T$. This time-delay phenomenon is reproduced
by the MSFA.

The MSFA predictions of the ionization velocity (the time difference of the ionization yield) are presented in Fig. 2(c) with considering
all contributions and contributions only related to $p_y<0$.
Here, the curve of $p_y<0$ shows  large amplitudes at times  earlier than $t=5.25T$ or $t=5.75T$, implying that
electrons born at the rising part of the fundamental field also contribute remarkably to $p_y<0$, different from the predictions of SFA and  CM.
Further analyses tell that the long trajectory relating to a rescattering process still dominates the contributions to $p_y<0$,
and the short one associated with direct ionization dominates the cases of $p_y>0$.
The ratio of short-trajectory contributions to long ones is about 0.62,
which agrees with  the MSFA symmetry degree, indicating that
the Coulomb effect induces a remarkable increase of  long-trajectory contributions (more than $20\%$ relative to the SFA predictions) to ATI.
Previous TDSE studies with a suitably chosen momentum-space analysis have also indicated this remarkable increase \cite{Bohan}.
Here, with MSFA, the time-resolved mechanism can be accessed.

In Fig. 2(d), we further present the distributions $F(\mathbf{p},t_0)\equiv F(t_{i},t_0)$. Here $t_{i}$ is the MSFA prediction of the
ionization time
related to the trajectory ($\mathbf{p},t_0$). One can observe that the distributions deviate upward from the diagonal line,
and this upward deviation is more remarkable for the time $t_0$ later than $5.25T$ or $5.75T$.
The results clearly show the Coulomb induced large time lag (longer than 100 attoseconds on average)
for the exiting time $t_0$ and the ionization time $t_i$.

\begin{figure}[t]
\begin{center}
{\includegraphics[width=8.5cm,height=6.6cm]{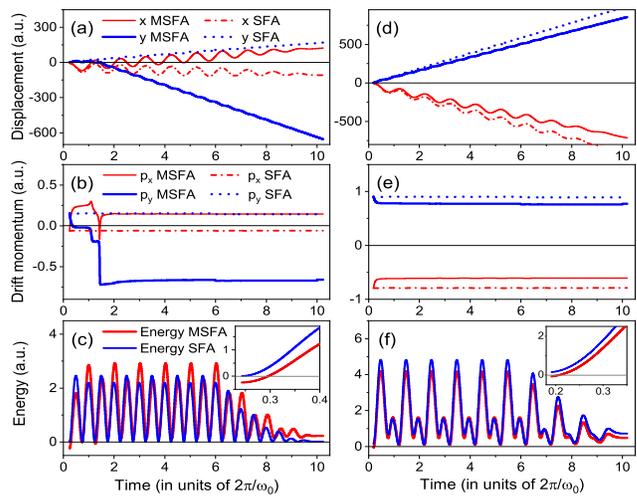}}
\caption{Comparisons for time evolution of displacements (a,d), momenta (b,e) and energy (c,f) of
two typical electron trajectories predicted by  MSFA and  SFA. Results in the left (right) column correspond
to the MSFA electron trajectory with (without) changing the  directions of its initial drift momenta under the influence of the Coulomb potential.
The insets in (c) and (f) show the enlarged results around the time origins of the trajectories in (c) and (f).}
\label{fig:graph1}
\end{center}
\end{figure}
\begin{figure}[t]
\begin{center}
{\includegraphics[width=8.5cm,height=6.6cm]{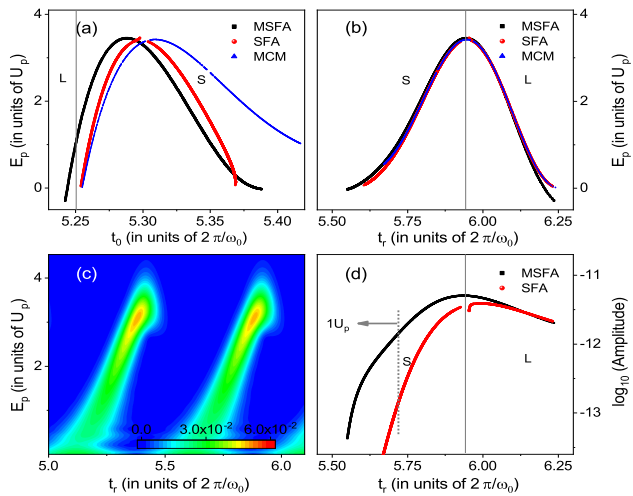}}
\caption{Comparisons of HHG long (L) and short (S) electron trajectories of He born at half a laser cycle,
obtained with MSFA, SFA, and MCM,
 for the exiting time $t_0$ (a) and the return time $t_r$ (b)  versus the return energy $E_p$ scaled with $U_p=E_0^2/(4\omega^2_0)$, and
the return time $t_r$
versus the harmonic amplitude (d). Results in (c) show time-frequency analyses of  TDSE simulations
where short-trajectory contributions are only considered \cite{Li2016}.
 The vertical-solid line in (a) indicates the peak time of $5.25T$.
Those in (b) and (d) indicate the HHG cutoff positions, which divide the trajectories into
long-trajectory (right branches) and short-trajectory (left) parts.}
\label{fig:graph1}
\end{center}
\end{figure}

Next, we perform analyses on how the Coulomb effect influences the electron trajectory.
In Fig. 3, we show two typical time-dependent MSFA electron trajectories with comparing with the SFA ones.
The SFA ones are obtained with assuming $V(\mathbf{r})\equiv0$ in the MSFA treatments.
We focus on the displacement of $\mathbf{r}(\mathbf{p},t)$, drift momentum of $\dot{\mathbf{r}}(\mathbf{p},t)-\mathbf{A}(t)$
and instantaneous energy of $E_a(t)$ for the electron trajectory.
Results in the left column show a SFA trajectory in the second quadrant with $p_x<0$ and $p_y>0$.
This trajectory corresponds to a short trajectory related to the fundamental field.
With considering the Coulomb effect, the MSFA predicts that this trajectory  will shift  to the fourth quadrant with $p_x>0$ and $p_y<0$.
The shifted trajectory corresponds to a long trajectory of the fundamental field, which is relating to a rescattering process,
as the displacement and the drift momentum of the shifted trajectory in Figs. 3(a) and 3(b) show.
According to the SFA, the energy of the trajectory is larger than zero just at the exiting time $t_0$ (i.e., the time origin of the trajectory).
By comparison, there is an obvious time difference for the MSFA ionization time $t_i$, at which the energy of the trajectory
becomes larger than zero, and the exiting time $t_0$, as the inset in Fig. 3(c) shows.
Results in the right column show the case of one electron trajectory for which the Coulomb effect is not remarkable.
Here, the MSFA and SFA predictions are similar for time evolution of this electron trajectory.
The Coulomb effect does not change the direction of the drift momentum, but modifies
the  values of displacement, momentum and energy for this trajectory. One can observe from the inset in Fig. 3(f),
the MSFA ionization time is also somewhat later than the SFA one.

\emph{Cases of HHG}.---
In the above discussions, we have shown that the Coulomb effect will remarkably influence ATI electron trajectories.
Due to that the ionization is the first step of many strong-field processes, one can expect that the Coulomb effect will also
 affect the dynamics of relevant processes.
As a case, in the following, we show that how the Coulomb effect influences  the HHG.

In Fig. 4, we show the comparisons of HHG electron trajectories of He in a single-color field (the fundamental $\omega_0$ field),
obtained with SFA, MSFA and
the modified classical model (MCM) that considers the exiting position \cite{chen2011}.
The time-frequency analysis \cite{Tong2000} of TDSE dipole acceleration  of He
with short-trajectory simulations  \cite{Li2016} is also shown here.
First, one can observe from Fig. 4(a), the exiting times of HHG electron trajectory predicted by these models differ remarkably
from each other. One of the remarkable differences is that the MSFA predicts some trajectories which begin at times  somewhat earlier than $5.25T$,
while the predictions of SFA and MCM for the exiting time are always later than $5.25T$, corresponding to the falling part of the field.
Moreover, in most of energy region, the MSFA predictions of the exiting time are earlier (about 25 attoseconds) than the SFA ones.
By comparison, the return times predicted by the models are nearer to each other in most of energy region,
especially for high energy, as seen in Fig. 4(b).

Remarkable differences are also observed for the HHG amplitudes predicted by MSFA and SFA, as seen in Fig. 4(d).
Here, the SFA amplitude is evaluated with the HHG saddle-point expression as in \cite{Lewenstein3}.
When the predictions of SFA and MSFA are similar for  long trajectories (right branches),
they differ remarkably for short trajectories (left).
The short-trajectory MSFA amplitudes are one order of magnitude higher than the SFA ones for lower energy (such as $E_p=U_p$,
indicated by the vertical-dotted line).
The MSFA short-trajectory results in Fig. 4(d) are nearer to the TDSE ones in Fig. 4(c), with showing comparable amplitudes
for lower and higher energy.
Our extended simulations show that these remarkable amplitude differences for MSFA and SFA hold for relatively
low laser intensities and short laser wavelengthes.
As seen in Fig. 4(a), the HHG exiting times predicted by MSFA
are nearer to the peak time of the laser field at which the tunneling amplitudes are larger.
This effect is mainly responsible for increased short-trajectory amplitudes of MSFA, in comparison with SFA.

In summary, we have studied the influence of Coulomb potential on temporal aspects of ATI and HHG.
We have shown that the Coulomb effect gives rise to a large time lag (longer than 100 attoseconds)
between the exiting time and the ionization time for ATI electron trajectories.
This time lag increases the contributions of long trajectory to ATI.
Accordingly, it also shifts the HHG exiting time towards the peak time of the laser field,
resulting in a remarkable increase of short-trajectory contributions  to HHG.
As the predictions of SFA or CM without considering the large time lag, are often used as references
to distill attosecond-resolved time information
from  photoelectron momentum distributions or  HHG spectra measured in experiments,
our work gives suggestions on relevant studies.
In addition,
our work also opens a new perspective for understanding
tunneling-triggered strong-field processes where the large time lag
is expected to play a nontrivial role.

This work is supported by the National Natural Science Foundation of China (Grant No. 91750111),
Research Team of Quantum Many-body Theorey and Quantum Control in Shaanxi Province (Grant No. 2017KCT-12),
and the Fundamental Research Funds for the Central Universities, China (Grant No. SNNU.GK201801009).

\end{document}